\documentclass[a4paper,12pt]{article}
\usepackage{amsmath}
\usepackage{amssymb}
\usepackage{fullpage}
\usepackage{verbatim}
\usepackage[T1]{fontenc}
\usepackage{hyperref}

\allowdisplaybreaks[1]

\renewcommand{\d}{\mathrm{d}}
\newcommand{\e}{\mathrm{e}}
\newcommand{\f}[2]{\frac{#1}{#2}}

\newcommand{\order}[1]{\mathcal{O}(#1)}
\newcommand{\tr}{\mathrm{tr}}

\begin{document}

\numberwithin{equation}{section}

\thispagestyle{empty}

\begin{flushright}
\small ITP-UH-08/12\\
\normalsize
\end{flushright}
\vspace{1cm}

\begin{center}

{\LARGE \bf On Cosmological Constants from}

\vskip3mm
 {\LARGE \bf $\alpha^{\prime}$-Corrections}

\vspace{1cm}
{\large Fri\dh rik Freyr Gautason, Daniel Junghans and Marco Zagermann}\\

\vspace{1cm}
{Institut f{\"u}r Theoretische Physik \&\\
Center for Quantum Engineering and Spacetime Research\\
Leibniz Universit{\"a}t Hannover, Appelstra{\ss}e 2, 30167 Hannover, Germany}\\

\vspace{1cm}
{\upshape\ttfamily fridrik.gautason, daniel.junghans, marco.zagermann@itp.uni-hannover.de}\\

\vspace{1cm}
\begin{abstract}
\noindent We examine to what extent perturbative $\alpha^{\prime}$-corrections can generate a small cosmological constant in warped string compactifications. Focusing on the heterotic string at lowest order in the string loop expansion, we show that, for a maximally symmetric spacetime, the $\alpha^{\prime}$-corrected 4D scalar potential has no effect on the cosmological constant. The only relevant terms are instead higher order products of 4D Riemann tensors, which, however, are found to vanish in the usual perturbative regime of the $\alpha^{\prime}$-expansion. The heterotic string therefore only allows for 4D Minkowski vacua to all orders in $\alpha^{\prime}$, unless one also introduces string loop and/or nonperturbative corrections or allows for curvatures or field strengths that are large in string units. In particular, we find that perturbative $\alpha^{\prime}$-effects cannot induce weakly curved $AdS_4$ solutions.

\end{abstract}

\end{center}

\newpage
\tableofcontents
\vspace{0.5cm}

\section{Introduction}

At sufficiently low energies and for small string coupling, perturbative string theory is well approximated by an effective two-derivative supergravity Lagrangian supplemented by small corrections coming from a double expansion in the slope parameter $\alpha^\prime$ and the string coupling $g_s$. The terms of the $\alpha^\prime$-expansion are higher derivative corrections to the supergravity action that account for the extended nature of the strings. They are negligible if the curvature of the background manifold and derivatives of the fields are small in units of $\alpha^{\prime}$. The terms coming from the $g_s$-expansion are loop corrections due to nontrivial topologies of the string world sheet, which are negligible in the semi-classical regime when the string coupling is small.

From a phenomenological point of view, such sub-leading corrections 
can have important consequences, as they may allow for solutions with properties that are forbidden at the two-derivative supergravity level. A well-known example in type IIB string theory are the AdS$_4$ solutions at large internal volume \cite{Balasubramanian:2005zx}, where $\alpha^\prime$-corrections \cite{Becker:2002nn} break the no-scale structure of the leading order Minkowski solutions found in  \cite{Giddings:2001yu} and contribute to a nonzero cosmological constant. In this example, however, the $\alpha^\prime$-corrections alone are not sufficient, and also non-perturbative quantum corrections from localized sources are needed in order to generate the AdS vacuum.

For the heterotic string, 
an analogous scenario was investigated in \cite{Anguelova:2010qd}, where the authors found that an interplay of the lowest order $\alpha^{\prime}$-correction \cite{Anguelova:2010ed} and non-perturbative 
effects could give rise to a similar large volume AdS vacuum in 4D, while the classical two-derivative supergravity action 
only admits Minkowski ground states.

In view of these constructions, one might wonder whether there could also be
situations where the perturbative $\alpha^{\prime}$-corrections alone already suffice
to generate a small non-vanishing cosmological constant in a controlled compactification scheme. 
This question should be easiest to study for the heterotic string, where D-branes and orientifold planes are absent, and the leading $\alpha^{\prime}$-corrections are completely known and already appear at
order $\mathcal{O}(\alpha^{\prime})$. Looking at the heterotic effective action at string tree-level, however, one might quickly conclude that $\alpha^{\prime}$-corrections alone can never suffice to generate vacua other than Minkowski space.
%
The apparent reason is that, in the absence of string loop or non-perturbative corrections, all terms in the heterotic effective action come from world sheets with spherical topology so that the action scales uniformly with the dilaton $\phi$:
\begin{equation}
S=\int \d^{10}x\sqrt{-g}\, \e^{-2\phi}\{\ldots\}
\end{equation}
(cf. \eqref{c}). As a consequence, the four-dimensional effective scalar potential likewise scales uniformly with the dilaton zero mode, and one would expect the 4D dilaton equation to be solved either if the potential vanishes on the solution or if there is a runaway to a free vacuum. It therefore seems obvious that heterotic string theory at string tree level can only lead to Minkowski solutions, and that a non-vanishing cosmological constant also requires string loop or non-perturbative quantum corrections. 
A related argument was employed by Dine and Seiberg in \cite{Dine:1985kv,Dine:1985he} to suggest that realistic string vacua might be strongly coupled.\footnote{A priori, all this also applies to the oriented closed string sector of the type II theories, but the inclusion of orientifold planes and D-branes leads to terms with a different dilaton scaling already at string tree-level. It is these different scalings that allow, e.g.,  for the classical AdS vacua of \cite{DeWolfe:2005uu} and that
have been exploited in attempts to construct ``classical''  de Sitter vacua in type II supergravity with (smeared) orientifold planes (see e.g. \cite{Haque:2008jz,Caviezel:2008tf,Flauger:2008ad}
for early discussions) that could evade the ``no-go'' theorems discussed in \cite{Hertzberg:2007wc,Silverstein:2007ac,Danielsson:2009ff,deCarlos:2009fq,Wrase:2010ew}.}

It is the purpose of this paper to re-address this question and in particular the seemingly trivial counter-argument against non-Minkowski vacua sketched in the previous paragraph. The reason is that the higher curvature terms among the
$\alpha^{\prime}$-corrections (e.g. the $\alpha^{\prime}\tr |R^+|^2$-terms in the heterotic string) also lead to contributions to the four-dimensional Einstein equation and the equations of motion for the moduli that involve higher powers of external Riemann tensors  and hence can \emph{not} be interpreted as a part of the effective scalar potential. It is therefore a priori not clear whether the scaling argument sketched above is still valid or whether nontrivial effects might emerge from such higher order terms.

That these effects exist follows from explicitly known AdS$_4$-compactifications of the heterotic string when the effective action is truncated after the lowest order $\alpha^{\prime}$-corrections (see e.g. \cite{Lechtenfeld:2010dr,Chatzistavrakidis:2012qb}). In these solutions, the 4D cosmological constant turns out large, $\Lambda \sim \frac{1}{\alpha^{\prime}}$, so that the effects of even higher $\alpha^{\prime}$-corrections are difficult to estimate offhand and would require more explicit calculations \cite{Lechtenfeld:2010dr}.

In an interesting recent paper \cite{Green:2011cn}, on the other hand, it was investigated whether the $\alpha^{\prime}$-corrections of the heterotic string could also give rise to a \emph{small} cosmological constant $\Lambda\sim \alpha^{\prime} C$, where $C$ is a 6D integral over fields such as the dilaton or the warp factor with four internal derivatives. Intriguingly, the authors found that de Sitter vacua of this type are excluded, but raised the possibility of warped AdS$_4$ compactifications as an $\mathcal{O}(\alpha^{\prime})$-effect.
Proving this requires only some of the 10D field equations, and it was left as an open problem to check whether really all field equations could be satisfied at the considered order in the $\alpha^{\prime}$-expansion. 

In this paper we investigate to what extent the usual scaling analysis of the 4D effective potential is invalidated by higher curvature terms in the $\alpha^{\prime}$-expansion and 
check whether this expansion can yield perturbatively small cosmological constants of order $\mathcal{O}(\alpha^{\prime})$ or higher.
The main result of our analysis is that this is in general not possible 
 at string tree-level. This follows from the four-dimensional Einstein equation and the dilaton equation, which can be combined to yield a constraint of the form 
\begin{equation}
\Lambda = \sum_{m,n} c_{mn} \alpha^{\prime m} \Lambda^n,  \qquad m,n>0,
\end{equation}
where $c_{mn}$ are numerical coefficients containing integrals over internal fields and their derivatives.  Assuming a perturbative $\alpha^{\prime}$-expansion for $\Lambda$, one then obtains $\Lambda = 0$ as the only solution to all orders in $\alpha^\prime$, as we will explain in more detail below.

This paper is organized as follows. In Section \ref{het} we establish our notation and detail a simple argument (cf. \cite{Held:2010az}) showing that heterotic supergravity with the first order $\alpha^\prime$-corrections does not yield solutions with a nonzero cosmological constant to that order. In Appendix \ref{lohet} we investigate the proposed warped AdS solutions of \cite{Green:2011cn} at order $\mathcal{O}(\alpha^{\prime})$ more directly, showing explicitly that the given $\mathcal{O}(\alpha^{\prime})$-expression for the cosmological constant is really of higher order and thus could compete with neglected terms in the action. In Section \ref{nogo} we then show how the argument of Section \ref{het}  can be extended to all orders in the $\alpha^\prime$-expansion and that it is completely independent of the details of the $\alpha^{\prime}$-corrected 4D scalar potential. We conclude with Section \ref{disc}, where we discuss several ways to circumvent this ``no-go theorem'', its relation to the Dine-Seiberg problem, and possible effects of violations of the effective potential description. Appendix \ref{riemann} collects some useful identities for the Riemann tensor, and Appendix \ref{10d} contains the ten-dimensional version of the argument of Section \ref{nogo}.

\section{A ``no-go theorem''}
\label{nogothm}
In this section, we discuss a simple argument showing that tree-level heterotic string theory with its first order $\alpha^\prime$-corrections does not have 4D de Sitter or anti-de Sitter vacua with a perturbatively small cosmological constant at this order \cite{Held:2010az}. We then show that the argument can in fact be extended to all orders in the $\alpha^\prime$-expansion. Our assumptions throughout the paper are as follows:
\begin{itemize}
\item We consider compactifications to four dimensions that respect maximal four-dimen-sional spacetime symmetry, i.e.:
\begin{itemize} 
\item The 10D metric is a warped product of a maximally symmetric 4D spacetime (parameterized by coordinates $x^\mu$; $\mu,\nu,\ldots=0,\ldots,3$) and a 6D compact manifold (parameterized by
$y^m$; $m,n,\ldots=4,\ldots,9$),
\begin{eqnarray}
\d s^2 &=& \e^{2A} \d s_4^2 + \d s_6^2,
\end{eqnarray}
where $\e^{2A}$ depends on the 6D coordinates only, and $\d s_4^2$ describes an unwarped 4D Minkowski, de Sitter or anti-de Sitter spacetime.
\item All 4D parts of tensor and spinor fields vanish (up to gauge choices) except for combinations that can be built from the 4D (unwarped) metric, its
Riemann tensor or its volume form. This means, in particular, that  there are no spacetime-filling fluxes\footnote{We express everything in terms of the Yang-Mills field strength $F$ and the NS 3-form $H$, which have a too small rank to be spacetime-filling in 4D. The Hodge duals
of purely 6D fluxes of these fields would of course generically have spacetime-filling components, but they do not appear explicitly in our formalism.} and that all 4D covariant derivatives 
of all tensor fields, including the dilaton and the Riemann tensor, can be set to zero on the solution.\footnote{Note that for maximally symmetric spaces, the Riemann tensor becomes an algebraic combination of metric tensors, and therefore its covariant derivative vanishes.} Furthermore, the Lorentz-Chern-Simons 3-form does not contribute to the equations of motion in maximally symmetric backgrounds \cite{Campbell:1990fu}.
\end{itemize}
\item String-loop and/or non-perturbative corrections to the action are disregarded.
\item $\alpha^{\prime}$ is a meaningful expansion parameter in the sense that all field variations are small over a string length and the $\alpha^{\prime}$-corrections can be organized in a perturbative expansion about the zero-slope limit.\footnote{The $\alpha^\prime$-expansion differs from the derivative expansion in that some terms appear at higher orders than suggested by the number of their derivatives. An example is the term $\tr |F|^2$ which, although a two derivative term, appears at $\order{\alpha'}$. It should be noted though that our analysis does not depend on which of the two expansion schemes is used.}
\end{itemize}

\subsection{Heterotic supergravity with leading $\alpha^\prime$-corrections}
\label{het}

In string frame, the heterotic supergravity action with leading $\alpha^\prime$-corrections reads
(for simplicity we set the 10D gravitational coupling $\kappa^2=\frac{1}{2}$)
\begin{equation}
S = \int \d ^{10} x \sqrt{-g}\, \e^{-2\phi} \left\{{ R + 4 (\partial \phi)^2 - \frac{1}{2} |H|^2 - \frac{\alpha^\prime}{4} \left[{\mathrm{tr} |F|^2 - \mathrm{tr} |R^+|^2 }\right]  } + \mathcal{O }(\alpha^{\prime 2})\right\}\label{c}
\end{equation}
with $|H|^2 = \f{1}{6} H_{MNL}H^{MNL}$, $\tr |F|^2 = \f{1}{2}\tr F_{MN}F^{MN}$ and $\tr |R^+|^2 = \f{1}{2} R^+_{MNPQ}R^{+MNPQ}$. Here, $\phi$ denotes the dilaton, $F$ is the Yang-Mills field strength, and $R^+_{MNP}\vphantom{}^Q$ is the Riemann tensor constructed from the torsionful connection $\Gamma^+\vphantom{}^M_{NL} = \Gamma^{M}_{NL} - \frac{1}{2} H_{NL}\vphantom{}^{M}$. $H$ is the $\alpha^{\prime}$-corrected 3-form field strength of the NS 2-form potential $B$, 
\begin{equation}
H = \d B + \f{\alpha'}{4}\left({\omega_{3 \text{L}} - \omega_{3 \text{Y}}}\right),
\end{equation}
where $\omega_{3 \text{L}}$ and $\omega_{3 \text{Y}}$ denote the Chern-Simons 3-forms formed from the spin connection and the Yang-Mills gauge field, respectively.

For our argument, it is sufficient to look at the field equations of the dilaton and the external metric.
This can be done either by using a 4D effective action approach or by working directly with the 10D field equations. We describe the 4D effective action approach here and sketch the analogous 10D argument 
in Appendix \ref{10d}.

For the 4D argument, we can restrict our attention to the zero mode, $\tau$, of the dilaton,
which we define by separating off  the higher Kaluza-Klein modes,
\begin{equation}
\e^{-\phi} = \tau \e^{-\phi_\textrm{KK}}. 
\end{equation}
Here $\phi_\textrm{KK}$ denotes the sum of all remaining KK-modes, which we integrate out by simply setting them equal to their on-shell values. It does not matter for our argument whether $\tau$ or one of the KK modes has the lowest mass (or whether they even combine with other degrees of freedom in the low energy EFT as suggested in \cite{Underwood:2010pm}) as can be seen directly from the equivalent ten-dimensional analysis in Appendix \ref{10d}.

On-shell, all fields in 4D must be covariantly constant by maximal symmetry, so we can henceforth ignore 
any $x^\mu$-dependence of $\tau$ and only need to keep track of  $\tau$ itself in the action, but not of its derivatives. 

The only other field whose dynamics we need to consider is the external metric $g_{\mu\nu}$.
Switching to four-dimensional Einstein frame, we  define a new 4D metric $\tilde{g}_{\mu\nu}$ by
\begin{equation}
 \tilde g_{\mu\nu} \equiv \mathcal{V} \tau^2 \e^{-2A} g_{\mu\nu}. \label{frame}
\end{equation}
Here
\begin{equation}
\mathcal{V} \equiv \int \d ^{6} y \sqrt{g_6}\, \e^{-2\phi_\textrm{KK}+2A},
\end{equation}
which can again be treated as constant in 4D by maximal symmetry.

Performing this rescaling, we then obtain an effective 4D action for $\tilde{g}_{\mu\nu}$ and $\tau$ of the form
\begin{align}
S = \int \d ^{4} x \sqrt{-\tilde g_4} \left\{{\tilde R_4 - V + W}\right\},
\end{align}
where we have split the action into the Einstein-Hilbert term and two extra contributions. $V$ contains all terms that are constructed from fields without external indices, whereas $W$ contains all terms that include fields with 4D  spacetime indices. In the absence of $W$, $V$ is just the usual effective potential.

Using (\ref{c}),
these two terms are given by 
\begin{align}
V = - & \int \d ^6 y \sqrt{g_6}\, \e^{-2\phi_\textrm{KK}+4A} \frac{1}{\tau^{2} \mathcal{V}^2} \notag \\ &{} \times \left[{ \vphantom{|R^{\textrm{(w)}}_{\mu\nu\lambda}\vphantom{}^\rho|^2} R_6 - 20 (\partial A)^2 - 8 \nabla^2 A + 4 (\partial \phi)^2 - \frac{1}{2} |H|^2 - \frac{\alpha^\prime}{4} \left({\mathrm{tr} |F|^2 - |R^+_6|^2}\right)}\right] +\mathcal{O}(\alpha^{\prime 2})
\end{align}
and
\begin{equation}
W = \int \d ^6 y \sqrt{g_6}\, \e^{-2\phi_\textrm{KK}} \left[{\frac{\alpha^\prime \tau^2}{4} |\tilde R_{\mu\nu\lambda}\vphantom{}^\rho|^2 - \frac{\alpha^\prime}{2\mathcal{V}} \e^{2A} \tilde R_4 (\partial A)^2 }\right] +\mathcal{O}(\alpha^{\prime 2}), \label{w}
\end{equation}
where we have evaluated the curvature terms $R$ and $\mathrm{tr} |R^+|^2$ for the tilded metric \eqref{frame} and expressed them in terms of $\tilde R_4$ and $|\tilde R_{\mu\nu\lambda}\vphantom{}^\rho|^2 = \frac{1}{2} \tilde R_{\mu\nu\lambda\rho}\tilde R^{\mu\nu\lambda\rho}$ as well as a term $|R^+_6|^2$ containing various internal fields. Further details and the definition of $|R^+_6|^2$ can be found in Appendix \ref{riemann}.

Using the scaling $V \sim \tau^{-2}$, one finds the 4D dilaton equation, 
\begin{equation}
2 V + \frac{\alpha^\prime \tau^2}{2} |\tilde R_{\mu\nu\lambda}\vphantom{}^\rho|^2 \int \d ^6 y \sqrt{g_6}\, \e^{-2\phi_\textrm{KK}} = 0, \label{ex1}
\end{equation}
and the trace of the four-dimensional Einstein equation,
\begin{equation}
\tilde R_4 - 2 V - \frac{\alpha^\prime}{2\mathcal{V}} \tilde R_4 \int \d ^6 y \sqrt{g_6}\, \e^{-2\phi_\textrm{KK}+2A}\, (\partial A)^2 = 0, \label{ex2}
\end{equation}
where we have neglected the variation with respect to the connection as it would give rise to covariant derivatives upon partial integration, which vanish due to maximal symmetry. Combining the two equations such that $V$ cancels out and substituting $\tilde R_{\mu\nu\lambda\rho} = \frac{2}{3} \Lambda \tilde g_{\lambda[\mu} \tilde g_{\nu]\rho}$ then yields an equation of the form
\begin{equation}
\Lambda = \alpha^\prime \left({c_{11} \Lambda + c_{12} \Lambda^2}\right) + \mathcal{O}(\alpha^{\prime 2}), \label{ex3}
\end{equation}
where $c_{11}$ and $c_{12}$ are given by
\begin{equation}
c_{11} = \f{1}{2\mathcal{V}}\int \d ^6 y \sqrt{g_6}\, \e^{-2\phi_\textrm{KK}+2A}\, (\partial A)^2,\qquad
c_{12} = - \f{\tau^2}{3} \int \d ^6 y \sqrt{g_6}\, \e^{-2\phi_\textrm{KK}}.
\end{equation}


Given our assumption that we are in the regime of validity of the perturbative $\alpha^\prime$-expansion, \eqref{ex3} must be solved order by order with an ansatz of the form
\begin{equation}
\Lambda = \Lambda_0 + \alpha^\prime \Lambda_1 + \mathcal{O}(\alpha^{\prime 2})
\end{equation}
for the cosmological constant, where $\Lambda_0$ denotes the solution of the leading order supergravity equations without $\alpha^{\prime}$-corrections, $\alpha^\prime \Lambda_1$ is a correction due to next-to-leading order terms in the $\alpha^\prime$-expansion, and so on. It is straightforward to see that plugging this ansatz into \eqref{ex3} yields
\begin{equation}
\Lambda = \mathcal{O}(\alpha^{\prime 2}) \label{lambda1}
\end{equation}
as the only solution. Thus, perturbative heterotic string theory does not yield solutions with a nonzero cosmological constant up to corrections of order $\mathcal{O}(\alpha^{\prime 2})$.

Let us now compare this to the result of \cite{Green:2011cn}, where it was suggested that warped AdS vacua might be allowed in heterotic string theory as an $\mathcal{O}(\alpha^\prime)$-effect, i.e., 
\begin{equation}
\Lambda = - \alpha^\prime C + \mathcal{O}(\alpha^{\prime 2}), \label{lambda2}
\end{equation}
where $C$ is a non-negative constant which is built from a sum of squares of internal fields integrated over the internal manifold. At first sight, this seems to contradict the above argument that solutions with nonzero cosmological constant are not allowed at order $\mathcal{O}(\alpha^{\prime 1})$. However, one can show directly by means of the supergravity equations of motion that the terms contained in $C$ actually vanish at the order considered here such that $C=\order{\alpha^\prime}$. The right hand sides of \eqref{lambda1} and \eqref{lambda2} are therefore equal up to corrections of order $\mathcal{O}(\alpha^{\prime 2})$. For convenience, we give the details in Appendix \ref{lohet}.
\\

\subsection{General argument}
\label{nogo}

Let us now generalize the above argument to the heterotic string with $\alpha^\prime$-corrections of
arbitrarily high order. The effective action for the massless fields then reads \begin{equation}
S = \int \d ^{10} x \sqrt{-g}\, \e^{-2\phi} \left\{ R + 4 (\partial \phi)^2 - \frac{1}{2} |H|^2 + \alpha^\prime \text{-corrections}\right\}, \label{action}
\end{equation}
where all terms scale identically with respect to the dilaton if we neglect string loop or non-perturbative corrections as initially stated. 

Rescaling the metric as in \eqref{frame}, we obtain the action in four-dimensional Einstein frame
\begin{align}
S = & \int \d ^{4} x \sqrt{-\tilde g_4} \left\{{ \tilde R_4 - V + W }\right\}.
\end{align}
As in the previous section, we have split the action into an Einstein-Hilbert term $\tilde R_4$, a term $V$ containing all terms that are constructed from fields without external spacetime indices, and a term $W$ containing everything else.

In the absence of string loop or non-perturbative corrections, all terms in $V$ scale again  as $V \sim \tau^{-2}$ such that the dilaton equation yields
\begin{equation}
2 V + \tau \partial_\tau W = 0. \label{dil}
\end{equation}
Taking the trace of the four-dimensional Einstein equation, we furthermore find
\begin{equation}
\tilde R_4 - 2 V - W^\prime = 0, \qquad W^\prime \equiv \frac{\tilde g^{\mu\nu}}{\sqrt{-\tilde g_4}} \frac{\delta}{\delta \tilde g^{\mu\nu}} \left({ \int \d ^{4} x  \sqrt{-\tilde g_4} \, W}\right), \label{einst}
\end{equation}
where, as indicated, $W^\prime$ denotes all terms that are due to the variation of $W$ with respect to the external metric. 

Combining the two equations \eqref{dil} and \eqref{einst}, we then find
\begin{equation}
\tilde R_4 = - \tau \partial_\tau W + W^\prime. \label{ff}
\end{equation}

Although an explicit expression for the right hand side of this equation is only known for the first few orders in the $\alpha^\prime$-expansion, the general structure is rather simple: it is a sum of positive powers of the cosmological constant with coefficients built from integrals over internal fields and their derivatives. 

To see this, recall that our assumption of maximal 4D spacetime symmetry
implies that only the metric, the epsilon tensor and the Riemann tensor are nontrivial, all with vanishing covariant derivative. Considering first
the metric variations of $W$ that come from variations of connections (either within covariant derivatives or curvature tensors or Lorentz-Chern-Simons forms), one sees that these variations do not contribute to the 
right hand side of (\ref{ff}), as they
 would lead to
terms with a total 4D covariant derivative, which vanish by assumption.  
The only contributions to $W^{\prime}$ are therefore from
variations of metric tensors that appear algebraically in $W$ or in the metric determinant. 
As there are no nontrivial contractions of just the epsilon tensor and/or the metric, all these
terms must contain at least one Riemann tensor.\footnote{Note that there is no constant term in $W$: a constant has no external spacetime indices and hence would be part of $V$, which however cancels out in (\ref{ff}).} Similar remarks also apply to the dilaton variation
of $W$, so that the 
right hand side of (\ref{ff}) is a sum of terms that each involves at least one Riemann tensor.
Because of $\tilde R_{\mu\nu\lambda\rho} = \frac{2}{3} \Lambda \tilde g_{\lambda[\mu} \tilde g_{\nu]\rho}$, these then translate into positive powers of the cosmological constant, as claimed.

Since at leading order the supergravity action does not contain any terms that depend on the Riemann tensor except for the Einstein-Hilbert term, the terms in $W$ and $W^\prime$ are of order $\mathcal{O}(\alpha^\prime)$ or higher. We can therefore schematically rewrite \eqref{ff} as
\begin{equation}
\Lambda = \sum_{m,n} c_{mn} \alpha^{\prime m} \Lambda^n, \qquad m,n > 0, 
\label{gg}
\end{equation}
with some numerical coefficients $c_{mn}$ that in general contain integrals over contractions of warp factor terms, internal field strengths and curvatures, and so on.

Assuming again the validity of a perturbative $\alpha^{\prime}$-expansion, we need to solve \eqref{gg} order by order with an ansatz of the form
\begin{equation}
\Lambda = \Lambda_0 + \alpha^\prime \Lambda_1 + \alpha^{\prime 2} \Lambda_2 + ... \label{gggg}
\end{equation}
as in Section \ref{het}. This yields
\begin{equation}
\Lambda = 0
\end{equation}
as the only solution to all orders in the perturbative  $\alpha^\prime$-expansion.\footnote{We might also try to solve \eqref{gg} without expanding $\Lambda$ as in \eqref{gggg}. Assuming that $\Lambda\ne 0$, we can then divide by $\Lambda$ to get $1 \le \sum |c_{mn}\alpha'^m\Lambda^{n-1}|$. But this is again a contradiction to the assumption made in the beginning of Section \ref{nogothm}.} Hence, heterotic string theory yields Minkowski spacetime as the only maximally symmetric solution to all orders in the perturbative $\alpha^\prime$-expansion, unless one introduces loop and/or non-perturbative corrections. In particular, we don't find $\alpha^{\prime}$-generated $AdS_4$ vacua
with perturbatively small curvatures to be possible.
\\

\section{Discussion}
\label{disc}

Let us now discuss several implications of our findings. In particular, we will discuss possibilities to evade our above no-go argument, its relation to the Dine-Seiberg problem and the violation of the effective potential description due to higher order corrections to the supergravity action.
\\

\subsection{Evading the no-go theorem}

In Section \ref{nogo}, we have shown that heterotic string compactifications at string tree-level yield 4D Minkowski spacetime as the only maximally symmetric solution to all orders in a perturbative $\alpha^\prime$-expansion, unless one violates one of our initial assumptions. Let us now discuss these possible violations and how they evade our argument.

\subsubsection*{Loop and non-perturbative corrections/extended sources}

An obvious possibility to circumvent the argument of Section \ref{nogo} is the inclusion of terms that scale differently with respect to the dilaton than the tree-level terms considered here.  Natural candidates are string loop or non-perturbative corrections e.g. from gaugino condensation \cite{Derendinger:1985kk,Dine:1985rz}.
With such terms turned on, the dilaton and Einstein equations read
\begin{equation}
- \tau\partial_\tau V + \tau \partial_\tau W = 0, \qquad \tilde R_4 - 2 V - W^\prime = 0 \label{eoms}
\end{equation}
and can in general not be combined such that $V$ cancels out. The right hand side of \eqref{gg} may then contain terms which are independent of $\Lambda$, making solutions other than $\Lambda = 0$ possible. It would be interesting to see whether including the first loop correction at order $\mathcal{O}(\alpha^{\prime 3}g_s)$ could allow for purely perturbative solutions with a non-zero cosmological constant for the heterotic string.  

A different dilaton scaling may also be introduced if one includes extended sources such as the various types of D-branes and orientifold planes in type II string theory. Being an open string tree-level action, the DBI action scales only with $\e^{-\phi}$ and so would in general also invalidate our argument.
 In fact, in type II string theory, a large number of compactification scenarios with a nonzero cosmological constant have been proposed using D-branes and orientifolds as well as non-perturbative quantum corrections starting with \cite{Kachru:2003aw}. Heterotic string theory, on the other hand, is much more limited in this respect, as it does not contain D-branes and O-planes but would require dealing with less common extended objects. 

\subsubsection*{Spacetime-filling fluxes}

Since spacetime-filling fluxes are in general not forbidden by maximal symmetry, they can be used to invalidate our argument around \eqref{ff}, where we explained that all terms in $W$ are contractions of Riemann tensors and must therefore contain factors of the cosmological constant. In heterotic string theory, there are no spacetime-filling fluxes if spacetime is assumed to be four-dimensional. Compactifying to three dimensions, however, allows for solutions with a nonzero cosmological constant, if spacetime components of $H$ are turned on (see e.g. \cite{Kunitomo:2009mx}). In type II string theory, spacetime-filling RR-fluxes may also be present in compactifications to four dimensions and may lead to solutions with a nonzero cosmological constant.

\subsubsection*{Large higher derivative terms}

Another way to circumvent our no-go theorem is to leave the perturbative regime of the $\alpha^\prime$-expansion and consider solutions for which higher derivative terms are not small in units of $\alpha^\prime$. A truncation of the action at a finite order is then in general not guaranteed to be a good approximation to the full theory, because  higher order terms are not automatically suppressed.\footnote{This does of course not rule out that the truncated action could still capture the
essential features of a solution or that the higher order terms happen to be small or even vanish in certain cases.} This problem does of course not apply when supergravity is studied in its own right instead of being considered the low energy effective field theory of string theory. 
In any case, allowing curvature and derivatives of the fields to be large in units of $\alpha^\prime$, it is indeed possible to construct solutions with a nonzero cosmological constant that is large in units of $\alpha^{\prime}$. A good example are the heterotic AdS compactifications studied in \cite{Lechtenfeld:2010dr}, which are solutions to the heterotic supergravity action with linear $\alpha^\prime$-corrections that feature a curvature of order $\mathcal{O}(\frac{1}{\alpha^\prime})$. By construction, our argument does not make statements in this regime. 

\subsubsection*{Breaking maximal symmetry}

Requiring spacetime to be maximally symmetric implies a very limited field content such that, in the absence of spacetime-filling fluxes, all terms showing up on the right hand side of \eqref{ff} contain contractions of spacetime components of the Riemann tensor.
All of these terms can then be rewritten as a power of $\Lambda$ times some numerical factor, regardless of how the Riemann tensors are contracted. As explained earlier, this property ensures that the higher derivative curvature terms on the right hand side of \eqref{ff} are much smaller than the Ricci scalar on the left hand side, leading to constraint \eqref{gg} and the conclusion that only Minkowski solutions are possible.

For spacetimes without maximal symmetry, however, this need not be the case. The presence of various (spacetime) tensor fields then leads to new terms in \eqref{ff} which can be of the same order as the 4D Ricci scalar and thus generate a nonzero cosmological constant. Furthermore, it is not guaranteed anymore that higher derivative curvature terms in \eqref{ff} are negligible, since whether they are much smaller than the 4D Ricci scalar or can compete with it depends on how they are contracted. This is due to the well-known fact that for general spaces the magnitude of individual components of the Riemann tensor and the Ricci scalar need not be the same, so that the Riemann tensor can have large components even when the Ricci scalar is very small. In heterotic string theory, the Ricci scalar may then compete, for example, with the $\alpha^\prime |\tilde R_{\mu\nu\lambda}\vphantom{}^\rho|^2$ term and thus become nonzero. 

This is also the reason why it is not possible to extend our analysis to make a statement about the curvature of the internal space. An exception are compactifications on maximally symmetric spaces such as the six-sphere, which can be ruled out using an argument along the lines of Section \ref{nogo}, unless there exist six-form fluxes filling internal space. Since this only concerns a very restricted class of compactifications, our discussion does unfortunately not add much to the discussion of \cite{Douglas:2010rt}, where it is suggested that higher derivative corrections (or strong warping, see also \cite{Blaback:2010sj}) could in principle support an everywhere negative internal Ricci scalar, which is difficult to realize otherwise. 
\\

\subsection{The Dine-Seiberg problem}

In \cite{Dine:1985kv,Dine:1985he}, Dine and Seiberg used the dilaton behavior of the effective 4D scalar potential in the weak coupling limit to argue that, unless the effective potential is identically zero, there must in general either be a runaway to the free vacuum or a minimum at strong coupling. Using an analogous scaling analysis for the universal volume modulus, one may argue for similar difficulties
regarding compactifications at large volume (cf. e.g. \cite{Denef:2008wq} for a recent discussion). Progress in moduli stabilization techniques have since then led to many interesting scenarios where an interplay
of various scalar potential contributions suggest the existence of weakly coupled minima at controllably large volumes. Still many of the difficulties and complexities one encounters in these attempts can be 
traced back to the issues pointed out in \cite{Dine:1985kv,Dine:1985he}.

The argument given in the present paper, although somewhat similar in its consequences,
differs from the argument of  \cite{Dine:1985kv,Dine:1985he} in several ways.
First of all we do not really use or discuss moduli stabilization. Nor do we trace the 
dependence of the scalar potential on the volume modulus. In fact, the detailed form of the scalar potential and its moduli dependence play no role for our argument (except that we exploit the overall dilaton scaling to eliminate the scalar potential completely from the equation of interest (eq. (\ref{ff}))). Instead, the only terms that matter for our argument are higher order products of 4D Riemann tensors, which did not play a role for the arguments in \cite{Dine:1985kv,Dine:1985he}. 

Moreover, 
it could have been the case that terms that appear to be of lower order in the $\alpha^{\prime}$-expansion 
compete with terms that are explicitly of higher order in $\alpha^{\prime}$ without that the perturbative $\alpha^{\prime}$-expansion breaks down. An example for this are the $|H|^2$ and $|F|^2$ terms appearing in the heterotic supergravity action or gradient terms of the warp factor or the dilaton. As reviewed in Appendix \ref{lohet}, they are forced to be zero by the leading order equations of motion, if our initially stated assumptions hold. Including $\alpha^\prime$-corrections to the action, however, the equations of motion are modified such that the above terms can become nonzero and thus compete with higher order terms in the $\alpha^\prime$-expansion. This could have postponed the emergence of a nontrivial cosmological constant to a higher order than suggested by \cite{Green:2011cn}. Our argument from section \ref{nogo}, however, 
shows that this can not happen at any order in $\alpha^{\prime}$, regardless of the scalar potential.

\subsection{Violation of effective scalar potential description}

The effective scalar potential description is a standard tool in effective field theory which is widely used in the moduli stabilization literature. For solutions yielding a maximally symmetric spacetime, the effective potential is usually expected to fulfill two assumptions:
\begin{itemize}
\item The equations of motion are satisfied at a point in moduli space which is an extremum of $V$.
\item The value of $V$ at this point is proportional to the cosmological constant.
\end{itemize}
These assumptions are true if the effective action can be written in the form
\begin{align}
S = & \int \d ^{4} x \sqrt{-\tilde g_4} \left\{{ \tilde R_4 - V }\right\},
\end{align}
where $\tilde R_4$ is the only term in the Lagrangian that depends on the external metric, and $V$ is the only term that depends on the moduli.

It is interesting to note that both assumptions are generically violated by higher order effects in the $\alpha^\prime$-expansion, unless the cosmological constant is zero. This follows from \eqref{eoms} which on-shell yields
\begin{equation}
\partial_\tau V \neq 0, \qquad V \not \sim \Lambda.
\end{equation}
Hence, the equations of motion are in general satisfied at a point in moduli space which is not an extremum of $V$. Moreover, $V$ is not proportional to the cosmological constant anymore. 
This effect is usually completely negligible when the cosmological constant is small. For inflation scenarios with a very high energy scale, these corrections might be more sizeable, but when they are,
the validity of the perturbative $\alpha^{\prime}$-expansion would also be less obvious.

\section*{Acknowledgements}
This work was supported by the German Research Foundation (DFG) within the Cluster of Excellence "QUEST".

\appendix

\section{Riemann tensor with warping}
\label{riemann}

Let us compute the components of the Riemann tensor for the warped spacetime
\begin{equation}
\d s^2 = \frac{\e^{2A}}{\mathcal{V}\tau^2}\, \tilde g_{\mu\nu} \d x^\mu \d x^\nu + \d s_6^2.
\end{equation}
In order to express the full Riemann tensor $R_{MRN} \vphantom{}^P$ in terms of the Riemann tensor $\tilde R_{MRN} \vphantom{}^P$ of the unwarped metric $\tilde g_{MN}$, we use the formula
\begin{equation}
R_{MRN} \vphantom{}^P = - \tilde \nabla_M \Gamma^P_{RN} + \tilde \nabla_R \Gamma^P_{MN} + \Gamma^S_{NM} \Gamma^P_{RS} - \Gamma^S_{NR} \Gamma^P_{MS} + \tilde R_{MRN} \vphantom{}^P, \label{a}
\end{equation}
where $M,N,\ldots=0,\ldots,9$ denote 10D spacetime indices, $\Gamma^M_{NP} = \tfrac{1}{2} g^{MR} ( \tilde \nabla_P g_{RN} + \tilde \nabla_N g_{RP} - \tilde \nabla_R g_{NP})$ and $\tilde \nabla_M$ is the covariant derivative associated with the unwarped metric $\tilde g_{MN}$ (see e.g. \cite{Wald:1984rg}). This yields
\begin{eqnarray}
& R_{\mu\nu\lambda} \vphantom{}^\rho = - 2\, \frac{\e^{2A}}{\mathcal{V}\tau^2}\, \tilde g_{\lambda[\mu} \delta_{\nu]}^\rho (\partial A)^2 + \tilde R_{\mu\nu\lambda} \vphantom{}^\rho, \qquad R_{i j k} \vphantom{}^l = \tilde R_{i j k} \vphantom{}^l, & \notag \\
& R_{\mu j \lambda} \vphantom{}^l = - \frac{\e^{2A}}{\mathcal{V}\tau^2}\, \tilde g_{\mu\lambda} \nabla_j \partial^l A - \frac{\e^{2A}}{\mathcal{V}\tau^2}\, \tilde g_{\mu\lambda} (\partial_j A) (\partial^l A). &
\end{eqnarray}
Assuming that $H$ has only internal components, it follows from \eqref{a} that introducing torsion with $\Gamma^M_{NL} \to \Gamma^M_{NL} - \tfrac{1}{2} H_{NL}\vphantom{}^M$ modifies the internal components $R_{i j k} \vphantom{}^l$ of the Riemann tensor and, in case of nontrivial warping, also some of the spacetime components
\begin{equation}
R^+_{\mu\nu\lambda} \vphantom{}^\rho = R_{\mu\nu\lambda} \vphantom{}^\rho, \qquad R^+_{\mu j \lambda} \vphantom{}^l = R_{\mu j \lambda} \vphantom{}^l - \tfrac{1}{2} \Gamma^m_{\mu\lambda} H_{jm}\vphantom{}^l,
\end{equation}
where $\Gamma^m_{\mu\lambda} = -  \frac{\e^{2A}}{\mathcal{V}\tau^2}\, \tilde g_{\mu\lambda} \nabla^m A$. We thus find
\begin{align}
R & = \mathcal{V}\tau^2 \e^{-2A} \tilde R_4 + \tilde R_6 - 20 (\partial A)^2 - 8 \nabla^2 A, \label{rs} \\
\mathrm{tr} |R^+|^2 & = \mathcal{V}^2 \tau^4 \e^{-4A} |\tilde R_{\mu\nu\lambda}\vphantom{}^\rho|^2 - 2 \mathcal{V}\tau^2 \e^{-2A} \tilde R_4 (\partial A)^2 + |R^+_6|^2, \label{b}
\end{align}
where $\mathrm{tr} |R^+|^2 = \frac{1}{2} R^+_{MNPQ} R^{+MNPQ}$ and $|\tilde R_{\mu\nu\lambda}\vphantom{}^\rho|^2 = \frac{1}{2} \tilde R_{\mu\nu\lambda\rho}\tilde R^{\mu\nu\lambda\rho}$. For convenience, we also introduced the shortcut notation $|R^+_6|^2 = 12 \left[{(\partial A)^2}\right]^2 + 4 |R^+_{\mu j\lambda}\vphantom{}^l|^2 + |R^+_{ikl}\vphantom{}^m|^2$ to subsume all terms which in the tilded frame only depend on internal fields.
\\

\section{Leading order constraints on heterotic supergravity}
\label{lohet}

In \cite{Green:2011cn}, it was suggested that heterotic supergravity with leading $\alpha^\prime$-corrections could have solutions with a cosmological constant of the form
\begin{equation}
\Lambda = - \alpha^\prime C + \mathcal{O}(\alpha^{\prime 2}), \label{ggg}
\end{equation}
where $C$ is a non-negative constant given by
\begin{align}
C = \frac{1}{2\mathcal{V^\prime}} \int \d^6 y \sqrt{\tilde g_6}\, \e^{6A-\frac{\phi}{2}} & \left\{{\vphantom{\frac{1}{2}} 3 \left[{(\partial \omega)^2}\right]^2 + 2 |(\partial_m \omega)(\partial_n \omega) - \tilde \nabla_m \partial_n \omega - \tilde g_{mn} (\partial \omega)^2|^2}\right. \notag \\ & \left.{+ \frac{1}{2} \e^{-4\omega} |H_{mn}\vphantom{}^l \partial_l \omega|^2}\right\} \label{const}
\end{align}
with $\mathcal{V^\prime} = \int \d^6 y \sqrt{\tilde g_6}\, \e^{8A}$ and $\omega = A + \frac{\phi}{4}$. We will now show explicitly, using arguments similar to \cite{Strominger:1986uh,deWit:1986xg,Maldacena:2000mw},
that all terms in $C$ vanish up to higher order $\alpha^\prime$-corrections due to the leading order equations of motion. The result of \cite{Green:2011cn} is therefore not in conflict with the argument given in Section \ref{het}.

To omit confusion, we will stick to the metric conventions of \cite{Green:2011cn} in this appendix, which differ from those used in the main text of our paper. The unwarped metric is then defined as $\tilde g_{MN} = \e^{-2A} g_{MN}$, where $g_{MN}$ is the usual ten-dimensional Einstein frame metric. In the following, terms are always contracted with the Einstein frame metric, except for tilded objects and all terms in \eqref{const}, which are contracted with the unwarped metric $\tilde g_{MN}$.

The leading order dilaton equation in Einstein frame reads
\begin{equation}
\nabla_M \partial^M \phi + \frac{1}{2} \e^{-\phi} |H|^2 = \mathcal{O}(\alpha^\prime).
\label{xyz}
\end{equation}
Assuming that the dilaton only depends on the internal coordinates, we can write $\nabla_M \partial^M \phi = \e^{-10A} \tilde \nabla_m \e^{8A} \tilde g^{mn} \partial_n \phi$ and integrate over internal space to find
\begin{equation}
\f{1}{2}\int \d^6 y \sqrt{\tilde g_6}\, \e^{10A-\phi} |H|^2 = \mathcal{O}(\alpha^\prime)
\end{equation}
and hence
\begin{equation}
\e^{10A-\phi} |H|^2 = \mathcal{O}(\alpha^\prime).
\end{equation}

The traced internal and spacetime components of the leading order Einstein equation then read
\begin{equation}
-R_4 -2 R_6 + (\partial \phi)^2 = \mathcal{O}(\alpha^\prime), \qquad -3 R_4 - 2 R_6 + (\partial \phi)^2 = \mathcal{O}(\alpha^\prime).
\end{equation}
Combining the two equations and rewriting $R_4$ in terms of the unwarped metric yields
\begin{equation}
R_4 = \e^{-2A} \tilde R_4 - \frac{1}{2} \e^{-10A} \tilde \nabla^2 \e^{8A} = \mathcal{O}(\alpha^\prime). \label{t}
\end{equation}
We can now integrate over internal space to find $\e^{8A} \tilde R_4 = \mathcal{O}(\alpha^\prime)$ which with \eqref{t} implies that $\tilde \nabla^2 \e^{8A} = \mathcal{O}(\alpha^\prime)$. Hence the warp factor is a constant up to $\alpha^\prime$-corrections. The dilaton equation \eqref{xyz} then reduces to $\e^{-2A} \tilde \nabla^2 \phi = \mathcal{O}(\alpha^\prime)$ and therefore also $\phi$ is a constant up to $\alpha^\prime$-corrections.

We have thus shown that two-derivative terms involving the warp factor or the dilaton are at least of order $\mathcal{O}(\alpha^\prime)$, which implies that the four-derivative terms appearing in \eqref{const} are of even higher order. It follows that $C=\mathcal{O}(\alpha^\prime)$, and hence \eqref{ggg} yields
\begin{equation}
\Lambda = \mathcal{O}(\alpha^{\prime 2}).
\end{equation}

\section{Ten-dimensional argument}
\label{10d}

The result $\Lambda = 0$ can also be derived directly from the ten-dimensional equations of motion. We write the ten-dimensional action (\ref{action}) in the form
\begin{equation}
S = \int \d^{10}x \sqrt{-g}\, \e^{-2\phi} L,
\end{equation}
where $L$ includes all string theory $\alpha'$-corrections to the ten-dimensional supergravity. We start by pulling out an overall warp factor $g_{MN} = e^{2\omega} \tilde g_{MN}$. We will later on relate $\omega$ to the warp factor $A$ used in the main text. Writing the action in terms of the tilded metric $\tilde g$, we get
\begin{equation}
S = \int \d^{10}x \sqrt{-\tilde g}\, \e^{8\omega}\e^{-2\phi}\tilde L,
\end{equation}
notice the warp factor dependence in the action. The leading order terms of $\tilde L$ are
\begin{equation}
\tilde L = \tilde R -18(\tilde \nabla^2\omega + 4(\partial \omega)^2) + 4(\partial \phi)^2 - \f{1}{2}e^{-4\omega}|H|^2 + \order{\alpha'}.
\end{equation}
Assuming that $\tilde L$ only depends on the derivatives of $\phi$, the dilaton equation is easily derived up to a total derivative,
\begin{equation}\label{eq:10ddilatoneq}
0 = \f{1}{\sqrt{-\tilde g}}\f{\delta S}{\delta \phi} = -2\e^{8\omega}\e^{-2\phi}\tilde L + \text{total derivative}.
\end{equation}
Using this, we can simplify the Einstein equation
\begin{equation}\label{eq:10deinsteineq}
0 = \f{1}{\sqrt{-\tilde g}}\f{\delta S}{\delta \tilde g^{MN}} = \e^{8\omega}\e^{-2\phi}E_{MN} - \f{1}{2}\tilde g_{MN} \e^{8\omega}\e^{-2\phi} \tilde L =  \e^{8\omega}\e^{-2\phi} E_{MN} + \tilde g_{MN}(\text{total derivative}).
\end{equation}
The tensor $E_{MN}$ is simply the variation of the Lagrangian $\tilde L$.

Now take the ten-dimensional manifold to be a direct product of a six-dimensional compact space and maximally symmetric spacetime. We also let $\omega=\phi/4 + A$ to switch to the unwarped Einstein frame. Since spacetime is assumed maximally symmetric, all external covariant derivatives vanish and the total derivative in \eqref{eq:10deinsteineq} is a total derivative in internal space. We therefore look at the integrated traced Einstein equation, where the total derivatives drop out. To complete our analysis it is then enough for us to show that, when both indices lie in spacetime, $E_{\mu\nu}$ is a sum of terms that contain a positive power of the external curvature tensor. The only covariant quantities with external indices are the metric, the curvature tensor and the epsilon tensor. Keeping this in mind, there are only three possibilities that give a non-vanishing contribution to $E_{\mu\nu}$:
\begin{itemize}
\item Terms where one or both of the free indices are that of a curvature tensor. These obviously carry a positive power of the Riemann tensor and we are done.
\item Second are terms where the free indices are that of a metric, $E_{\mu\nu}\sim g_{\mu\nu}B$, coming from the ten-dimensional term $g_{MN}B$, where $B$ is a ten-dimensional scalar. This could be problematic, since $B$ does not have to involve the Riemann tensor. Clearly, these terms cannot occur as a result of varying the determinant, we already got rid of those using the dilaton equation. However we could have such terms from varying curvature tensors or covariant derivatives or more generally the connection. But varying the connection always gives a total derivative because of the equation
\begin{equation}
\delta \Gamma_{MN}^R = \f{1}{2}g^{RS}\left(\nabla_M \delta g_{NS} + \nabla_N \delta g_{MS} - \nabla_S \delta g_{MN}\right),
\end{equation}
and we see that $B$ must be a total derivative. Again this reduces to a total derivative in internal space and upon integration drops out.
\item The final possibility are terms where both external indices come from epsilon symbols. Clearly, an epsilon symbol must have four spacetime indices, and these must contract with something, the only possibility is a curvature tensor.
\end{itemize}
Other terms of the tensor $E_{MN}$ will be those, where the free indices are that of derivatives or fluxes etc. These all vanish in the maximally symmetric external spacetime. We have thus shown that all terms in the Einstein equation, traced with the external metric and integrated over internal space, contain a positive power of the Riemann tensor. This eventually leads to \eqref{gg}, and our result follows.
\\

\bibliographystyle{utphys}
\bibliography{groups}

\providecommand{\href}[2]{#2}\begingroup\raggedright\begin{thebibliography}{10}

\bibitem{Balasubramanian:2005zx}
V.~Balasubramanian, P.~Berglund, J.~P. Conlon, and F.~Quevedo, ``{Systematics
  of Moduli Stabilisation in Calabi-Yau Flux Compactifications},''
  \href{http://dx.doi.org/10.1088/1126-6708/2005/03/007}{{\em JHEP} {\bfseries
  03} (2005) 007},
\href{http://arxiv.org/abs/hep-th/0502058}{{\ttfamily arXiv:hep-th/0502058}}.

\bibitem{Becker:2002nn}
K.~Becker, M.~Becker, M.~Haack, and J.~Louis, ``{Supersymmetry breaking and
  alpha-prime corrections to flux induced potentials},'' {\em JHEP} {\bfseries
  0206} (2002) 060,
\href{http://arxiv.org/abs/hep-th/0204254}{{\ttfamily arXiv:hep-th/0204254
  [hep-th]}}.

\bibitem{Giddings:2001yu}
S.~B. Giddings, S.~Kachru, and J.~Polchinski, ``{Hierarchies from fluxes in
  string compactifications},''
  \href{http://dx.doi.org/10.1103/PhysRevD.66.106006}{{\em Phys. Rev.}
  {\bfseries D66} (2002) 106006},
\href{http://arxiv.org/abs/hep-th/0105097}{{\ttfamily arXiv:hep-th/0105097}}.

\bibitem{Anguelova:2010qd}
L.~Anguelova and C.~Quigley, ``{Quantum Corrections to Heterotic Moduli
  Potentials},'' \href{http://dx.doi.org/10.1007/JHEP02(2011)113}{{\em JHEP}
  {\bfseries 1102} (2011) 113},
\href{http://arxiv.org/abs/1007.5047}{{\ttfamily arXiv:1007.5047 [hep-th]}}.

\bibitem{Anguelova:2010ed}
L.~Anguelova, C.~Quigley, and S.~Sethi, ``{The Leading Quantum Corrections to
  Stringy Kahler Potentials},''
  \href{http://dx.doi.org/10.1007/JHEP10(2010)065}{{\em JHEP} {\bfseries 1010}
  (2010) 065},
\href{http://arxiv.org/abs/1007.4793}{{\ttfamily arXiv:1007.4793 [hep-th]}}.

\bibitem{Dine:1985kv}
M.~Dine and N.~Seiberg, ``{Couplings and Scales in Superstring Models},''
\href{http://dx.doi.org/10.1103/PhysRevLett.55.366}{{\em Phys. Rev. Lett.}
  {\bfseries 55} (1985) 366}.

\bibitem{Dine:1985he}
M.~Dine and N.~Seiberg, ``{Is the Superstring Weakly Coupled?},''
\href{http://dx.doi.org/10.1016/0370-2693(85)90927-X}{{\em Phys. Lett.}
  {\bfseries B162} (1985) 299}.

\bibitem{DeWolfe:2005uu}
O.~DeWolfe, A.~Giryavets, S.~Kachru, and W.~Taylor, ``{Type IIA moduli
  stabilization},'' \href{http://dx.doi.org/10.1088/1126-6708/2005/07/066}{{\em
  JHEP} {\bfseries 0507} (2005) 066},
\href{http://arxiv.org/abs/hep-th/0505160}{{\ttfamily arXiv:hep-th/0505160
  [hep-th]}}.

\bibitem{Haque:2008jz}
S.~S. Haque, G.~Shiu, B.~Underwood, and T.~Van~Riet, ``{Minimal simple de
  Sitter solutions},'' \href{http://dx.doi.org/10.1103/PhysRevD.79.086005}{{\em
  Phys.Rev.} {\bfseries D79} (2009) 086005},
  \href{http://arxiv.org/abs/0810.5328}{{\ttfamily arXiv:0810.5328 [hep-th]}}.
13 pages.

\bibitem{Caviezel:2008tf}
C.~Caviezel, P.~Koerber, S.~Kors, D.~L{\" u}st, T.~Wrase, {\em et~al.}, ``{On
  the Cosmology of Type IIA Compactifications on SU(3)-structure Manifolds},''
  \href{http://dx.doi.org/10.1088/1126-6708/2009/04/010}{{\em JHEP} {\bfseries
  0904} (2009) 010},
\href{http://arxiv.org/abs/0812.3551}{{\ttfamily arXiv:0812.3551 [hep-th]}}.

\bibitem{Flauger:2008ad}
R.~Flauger, S.~Paban, D.~Robbins, and T.~Wrase, ``{Searching for slow-roll
  moduli inflation in massive type IIA supergravity with metric fluxes},''
  \href{http://dx.doi.org/10.1103/PhysRevD.79.086011}{{\em Phys.Rev.}
  {\bfseries D79} (2009) 086011},
\href{http://arxiv.org/abs/0812.3886}{{\ttfamily arXiv:0812.3886 [hep-th]}}.

\bibitem{Hertzberg:2007wc}
M.~P. Hertzberg, S.~Kachru, W.~Taylor, and M.~Tegmark, ``{Inflationary
  Constraints on Type IIA String Theory},''
  \href{http://dx.doi.org/10.1088/1126-6708/2007/12/095}{{\em JHEP} {\bfseries
  12} (2007) 095},
\href{http://arxiv.org/abs/0711.2512}{{\ttfamily arXiv:0711.2512 [hep-th]}}.

\bibitem{Silverstein:2007ac}
E.~Silverstein, ``{Simple de Sitter Solutions},''
  \href{http://dx.doi.org/10.1103/PhysRevD.77.106006}{{\em Phys.Rev.}
  {\bfseries D77} (2008) 106006},
\href{http://arxiv.org/abs/0712.1196}{{\ttfamily arXiv:0712.1196 [hep-th]}}.

\bibitem{Danielsson:2009ff}
U.~H. Danielsson, S.~S. Haque, G.~Shiu, and T.~Van~Riet, ``{Towards Classical
  de Sitter Solutions in String Theory},''
  \href{http://dx.doi.org/10.1088/1126-6708/2009/09/114}{{\em JHEP} {\bfseries
  0909} (2009) 114},
\href{http://arxiv.org/abs/0907.2041}{{\ttfamily arXiv:0907.2041 [hep-th]}}.

\bibitem{deCarlos:2009fq}
B.~de~Carlos, A.~Guarino, and J.~M. Moreno, ``{Flux moduli stabilisation,
  Supergravity algebras and no-go theorems},''
  \href{http://dx.doi.org/10.1007/JHEP01(2010)012}{{\em JHEP} {\bfseries 1001}
  (2010) 012},
\href{http://arxiv.org/abs/0907.5580}{{\ttfamily arXiv:0907.5580 [hep-th]}}.

\bibitem{Wrase:2010ew}
T.~Wrase and M.~Zagermann, ``{On Classical de Sitter Vacua in String Theory},''
  \href{http://dx.doi.org/10.1002/prop.201000053}{{\em Fortschr. Phys.}
  {\bfseries 58} (2010) 906--910},
\href{http://arxiv.org/abs/1003.0029}{{\ttfamily arXiv:1003.0029 [hep-th]}}.

\bibitem{Lechtenfeld:2010dr}
O.~Lechtenfeld, C.~N{\" o}lle, and A.~D. Popov, ``{Heterotic compactifications
  on nearly K{\" a}hler manifolds},''
  \href{http://dx.doi.org/10.1007/JHEP09(2010)074}{{\em JHEP} {\bfseries 1009}
  (2010) 074},
\href{http://arxiv.org/abs/1007.0236}{{\ttfamily arXiv:1007.0236 [hep-th]}}.

\bibitem{Chatzistavrakidis:2012qb}
A.~Chatzistavrakidis, O.~Lechtenfeld, and A.~D. Popov, ``{Nearly K\'ahler
  heterotic compactifications with fermion condensates},''
\href{http://arxiv.org/abs/1202.1278}{{\ttfamily arXiv:1202.1278 [hep-th]}}.

\bibitem{Green:2011cn}
S.~R. Green, E.~J. Martinec, C.~Quigley, and S.~Sethi, ``{Constraints on String
  Cosmology},''
\href{http://arxiv.org/abs/1110.0545}{{\ttfamily arXiv:1110.0545 [hep-th]}}.

\bibitem{Held:2010az}
J.~Held, D.~L{\" u}st, F.~Marchesano, and L.~Martucci, ``{DWSB in heterotic
  flux compactifications},''
  \href{http://dx.doi.org/10.1007/JHEP06(2010)090}{{\em JHEP} {\bfseries 1006}
  (2010) 090},
\href{http://arxiv.org/abs/1004.0867}{{\ttfamily arXiv:1004.0867 [hep-th]}}.

\bibitem{Campbell:1990fu}
B.~A. Campbell, M.~J. Duncan, N.~Kaloper, and K.~A. Olive, ``{Gravitational
  dynamics with Lorentz Chern-Simons terms},''
\href{http://dx.doi.org/10.1016/S0550-3213(05)80045-8}{{\em Nucl. Phys.}
  {\bfseries B351} (1991) 778--792}.

\bibitem{Underwood:2010pm}
B.~Underwood, ``{A Breathing Mode for Warped Compactifications},''
  \href{http://dx.doi.org/10.1088/0264-9381/28/19/195013}{{\em Class. Quant.
  Grav.} {\bfseries 28} (2011) 195013},
\href{http://arxiv.org/abs/1009.4200}{{\ttfamily arXiv:1009.4200 [hep-th]}}.

\bibitem{Derendinger:1985kk}
J.~Derendinger, L.~E. Ibanez, and H.~P. Nilles, ``{On the Low-Energy d = 4, N=1
  Supergravity Theory Extracted from the d = 10, N=1 Superstring},''
\href{http://dx.doi.org/10.1016/0370-2693(85)91033-0}{{\em Phys.Lett.}
  {\bfseries B155} (1985) 65}.

\bibitem{Dine:1985rz}
M.~Dine, R.~Rohm, N.~Seiberg, and E.~Witten, ``{Gluino Condensation in
  Superstring Models},''
\href{http://dx.doi.org/10.1016/0370-2693(85)91354-1}{{\em Phys.Lett.}
  {\bfseries B156} (1985) 55}.

\bibitem{Kachru:2003aw}
S.~Kachru, R.~Kallosh, A.~D. Linde, and S.~P. Trivedi, ``{De Sitter vacua in
  string theory},'' \href{http://dx.doi.org/10.1103/PhysRevD.68.046005}{{\em
  Phys.Rev.} {\bfseries D68} (2003) 046005},
\href{http://arxiv.org/abs/hep-th/0301240}{{\ttfamily arXiv:hep-th/0301240
  [hep-th]}}.

\bibitem{Kunitomo:2009mx}
H.~Kunitomo and M.~Ohta, ``{Supersymmetric AdS$_3$ solutions in Heterotic
  Supergravity},'' \href{http://dx.doi.org/10.1143/PTP.122.631}{{\em Prog.
  Theor. Phys.} {\bfseries 122} (2009) 631--657},
\href{http://arxiv.org/abs/0902.0655}{{\ttfamily arXiv:0902.0655 [hep-th]}}.

\bibitem{Douglas:2010rt}
M.~R. Douglas and R.~Kallosh, ``{Compactification on negatively curved
  manifolds},'' \href{http://dx.doi.org/10.1007/JHEP06(2010)004}{{\em JHEP}
  {\bfseries 1006} (2010) 004},
\href{http://arxiv.org/abs/1001.4008}{{\ttfamily arXiv:1001.4008 [hep-th]}}.

\bibitem{Blaback:2010sj}
J.~Blaback, U.~H. Danielsson, D.~Junghans, T.~Van~Riet, T.~Wrase, and
  M.~Zagermann, ``{Smeared versus localised sources in flux
  compactifications},'' \href{http://dx.doi.org/10.1007/JHEP12(2010)043}{{\em
  JHEP} {\bfseries 1012} (2010) 043},
\href{http://arxiv.org/abs/1009.1877}{{\ttfamily arXiv:1009.1877 [hep-th]}}.

\bibitem{Denef:2008wq}
F.~Denef, ``{Les Houches Lectures on Constructing String Vacua},''
\href{http://arxiv.org/abs/0803.1194}{{\ttfamily arXiv:0803.1194 [hep-th]}}.

\bibitem{Wald:1984rg}
R.~M. Wald, ``{General Relativity},''.
The University of Chicago Press, 1984.

\bibitem{Strominger:1986uh}
A.~Strominger, ``{Superstrings with Torsion},''
\href{http://dx.doi.org/10.1016/0550-3213(86)90286-5}{{\em Nucl.Phys.}
  {\bfseries B274} (1986) 253}.

\bibitem{deWit:1986xg}
B.~de~Wit, D.~Smit, and N.~Hari~Dass, ``{Residual Supersymmetry of Compactified
  D=10 Supergravity},''
\href{http://dx.doi.org/10.1016/0550-3213(87)90267-7}{{\em Nucl.Phys.}
  {\bfseries B283} (1987) 165}.

\bibitem{Maldacena:2000mw}
J.~M. Maldacena and C.~Nunez, ``{Supergravity description of field theories on
  curved manifolds and a no go theorem},''
  \href{http://dx.doi.org/10.1142/S0217751X01003937}{{\em Int.J.Mod.Phys.}
  {\bfseries A16} (2001) 822--855},
\href{http://arxiv.org/abs/hep-th/0007018}{{\ttfamily arXiv:hep-th/0007018
  [hep-th]}}.

\end{thebibliography}\endgroup

\end{document}